\documentclass[twocolumn,floatfix,prl,aps,reprint]{revtex4-1}
\usepackage{float}
\usepackage{graphicx}
\usepackage{amsmath}
\usepackage{bm}
\usepackage{color}
\newcommand{\sgn}{\text{sgn}}
\usepackage[colorlinks=true, letterpaper=true, pdfstartview=FitV, linkcolor=blue, citecolor=blue, urlcolor=blue]{hyperref}
\begin{document}
\title{Time-Reversal-Invariant $Z_4$ Fractional Josephson Effect
}
\author{Fan Zhang and C. L. Kane}
\affiliation{Department of Physics and Astronomy, University of Pennsylvania, Philadelphia, PA 19104, USA}
\begin{abstract}
We study the Josephson junction mediated by the quantum spin Hall edge states
and show that electron-electron interactions lead to a dissipationless fractional Josephson effect in the presence of time-reversal symmetry.
Surprisingly, the periodicity is $8\pi$, corresponding to a Josephson frequency $eV/2\hbar$.
We estimate the magnitude of interaction induced many-body level splitting responsible for this effect
and argue that it can be measured using tunneling spectroscopy.
For strong interactions we show that the Josephson effect is associated with
the weak tunneling of charge $e/2$ quasiparticles between the superconductors.
Our theory describes a fourfold ground state degeneracy that is similar to that of coupled ``fractional'' Majorana
modes, but is protected by time reversal symmetry.
\end{abstract}
\maketitle

Topological Superconductivity is a topic of current interest because of its potential for providing a method for storing and manipulating quantum information~\cite{kitaev,review1,review2}.   The simplest implementation of this proposal uses the Majorana modes predicted to occur at the end of a 1D topological superconductor.
A promising route to achieve this is to employ a proximity effect device that combines an ordinary superconductor with a material that has a single helical band~\cite{fk1,fk2,nilsson,sau,alicea,lutchyn,oreg}.
There has been progress towards this goal using InSb quantum wires~\cite{mourik} and
using the edge states of quantum spin Hall (QSH) insulators in HgTe~\cite{yacoby} and InAs/GaSb~\cite{du} quantum wells.

One of the most basic consequences of topological superconductivity is the fractional Josephson effect~\cite{kitaev,kwon,fk2,z2z2}.
This occurs due to the coherent tunneling of electrons between the Majorana end states of two 1D topological superconductors.
A pair of Majorana modes defines two states that are split by the electron tunneling and are distinguished by their local fermion parity.
Advancing the phase difference $\phi$ across the junction by $2\pi$ interchanges the two states,
which leads to a $4\pi$ periodicity for each state as an adiabatic function of $\phi$.
This resembles a ``$Z_2$ pump''~\cite{fkz2pump}, and can be referred to as a ``$Z_2$ fractional Josephson effect''.
It gives rise to an AC Josephson effect with half the conventional Josephson frequency,
provided scattering from thermally excited bulk quasiparticles is sufficiently suppressed.

The fractional Josephson effect was originally proposed by Kitaev\cite{kitaev} using a model 1D spinless p wave superconductor.
It was later found that similar physics can arise for a Josephson junction mediated by the QSH edge states~\cite{fk2}.
In this case, a weakly coupled junction is formed by introducing a magnetic gap to the QSH edge states between the superconductors,
creating two weakly coupled Majorana modes at the superconductor-magnet interfaces.
For this construction it was essential that the time-reversal symmetry (TRS) be explicitly broken in the junction region to produce a dissipationless Josephson effect.
If it is not, then the Andreev bound states do not decouple from the bulk states,
and bulk quasiparticles are necessarily generated as $\phi$ is adiabatically advanced.

\begin{figure}[b!]
\scalebox{0.46}{\includegraphics*{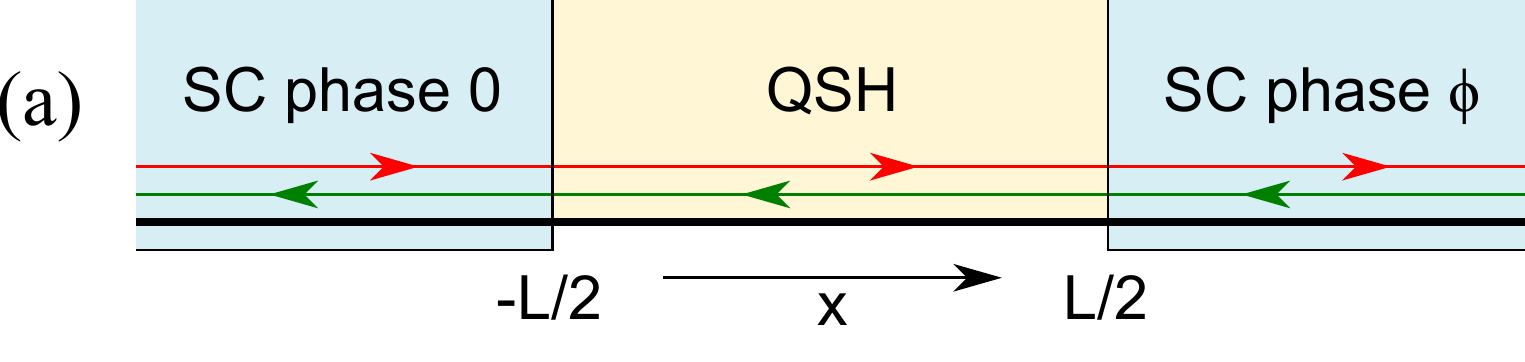}}
\scalebox{0.41}{\includegraphics*{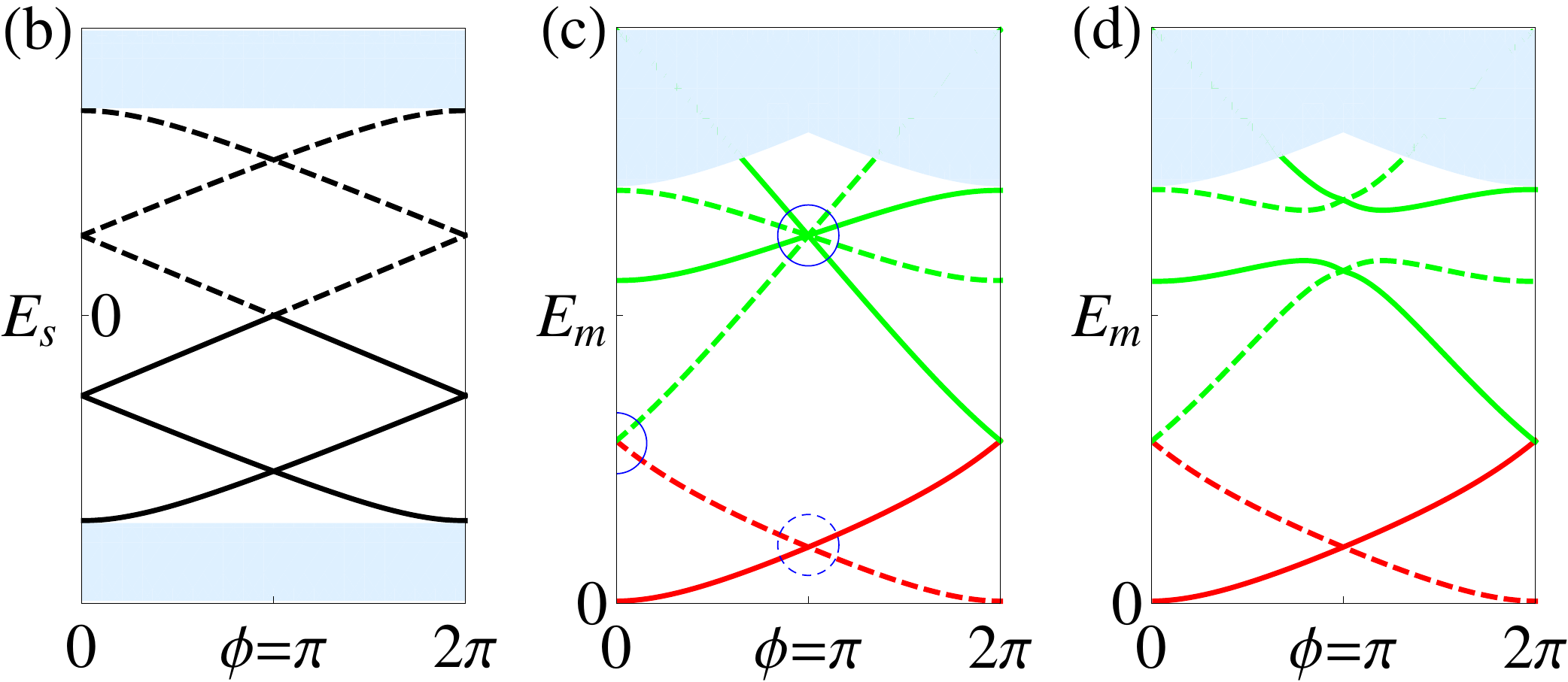}}
\caption{(a) Schematic of a Josephson junction mediated by the QSH edge states.
(b) The single-particle BdG spectrum of the junction as a function of $\phi$,
with Kramers degeneracies at $\phi = 0$ and $\pi$.
(c) The spectrum of many-body states corresponding to (b), including a fourfold degeneracy at $\phi=\pi$.
(d) The fourfold degeneracy is lifted by electron-electron interactions,
leading to an $8\pi$ periodicity of the four lowest states.}
\label{fig1}
\end{figure}

In this paper we will show that electron-electron interactions restore the time-reversal-invariant fractional Josephson effect,
but lead to an $8\pi$ periodicity of the Josephson current, which we refer to as a ``$Z_4$ fractional Josephson effect''.
We estimate the magnitude of the interaction induced splitting in the many-body excitation spectrum and
propose
a method for detecting this effect using tunneling spectroscopy.
When the QSH edge states between the superconductors are ungapped, the junction is necessarily strongly coupled.
However, if the edge states acquire a gap, we show that the $Z_4$ Josephson effect is associated with the tunneling of charge $e/2$ quasiparticles.
In this weak coupling limit the junction has a fourfold ground state degeneracy that is lifted with a characteristic pattern by tunneling.
The interface between the gapped edge states and the superconductor exhibits a domain wall excitation,
which is analogous to the fractional Majorana mode~\cite{clarke,lindner,cheng,vaezi,mong,loss} and related to a $Z_4$ parafermion.
However, there are also important differences with the parafermion theory, which we will clarify.

We begin with the model for a Josephson junction at the edge of a QSH insulator~\cite{fk2}, described by the Bogoliubov de Gennes (BdG) Hamiltonian
\begin{equation}
{\cal H}_{BdG} = \tau^z(-i \hbar v_F\sigma^z \partial_x-\mu) + \Delta_1(x) \tau^x + \Delta_2(x) \tau^y\,,
\label{h0}
\end{equation}
where $\vec\sigma$ ($\vec\tau$) are Pauli matrices in spin (particle-hole) space and $\Delta = \Delta_1 + i\Delta_2$ is the proximity induced pair potential.
We suppose that $\Delta(x<-L/2) = \Delta_0$, $\Delta(x>L/2) = \Delta_0 e^{i\phi}$, and $\Delta(|x|<L/2) = 0$.
The single-particle spectrum is shown in Fig.~\ref{fig1}(b).
The Andreev bound states at the phase difference $\phi=0$ and $\pi$ are necessarily Kramers degenerate.
This leads to a breakdown of the AC Josephson effect because as $\phi$ advances quasiparticles pass through the Kramers degeneracies and end up above the bulk gap leading to dissipation.

To go beyond the model where the junction is non-interacting, we consider in Fig.~\ref{fig1}(c) the many-body spectrum associated with Fig.~\ref{fig1}(b). The lowest state corresponds to the many-body ground state with all positive (negative) energy single-particle states in Fig.~\ref{fig1}(b) empty (occupied), whereas higher states are excitations with one or more quasiparticles excited. The local fermion parity of each state is indicated by the solid and dashed lines, and due to the fermion parity anomaly their identity switches when $\phi$ advances by $2\pi$. Fig.~\ref{fig1}(c) exhibits several degeneracies. The twofold degeneracies at $\phi=0$ and $\pi$ are Kramers degeneracies, protected by TRS. The twofold degeneracy at $\phi=\pi$ labeled by an open circle is even robust against TRS breaking since it involves two states with opposite fermion parity.
The fourfold degeneracy at $k=\pi$ labeled by a solid circle reflects the degeneracies of both $E=0$ and $E\neq0$ single-particle states.
However, this fourfold degeneracy is an artifact of the non-interacting electron approximation.
In the presence of electron-electron interactions it splits into two Kramers doublets,
each of which has two many-body states with opposite fermion parity.
There are thus only four low-energy states that mix among themselves as $\phi$ is adiabatically advanced, as indicated in Fig.~\ref{fig1}(d). Starting from the ground state at $\phi=0$, it takes four cycles to return to the original ground state, leading to an $8\pi$ periodicity in the current phase relation.
In the presence of a bias voltage $V$, this leads to an AC Josephson effect with a fundamental frequency $\omega_J = eV/2\hbar$, i.e., one quarter of the conventional Josephson frequency.

To establish the splitting at $\phi=\pi$ and estimate its magnitude we introduce a model
${\cal H} = {\cal H}_0 + {\cal H}_{I}$, where ${\cal H}_0$ is a second quantized version of Eq.~(\ref{h0}) and
\begin{equation}
{\cal H}_{I} = \lambda \int_{-L/2}^{L/2} n(x)^2\,.
\label{interaction}
\end{equation}
Here
$n(x) = \sum_\sigma c^\dagger_\sigma c_\sigma$ is the charge density.
We focus on the four degenerate excited states at $\phi = \pi$ and evaluate the splitting to first order in $\lambda$.
To proceed, we now determine the wavefunctions of single-particle Andreev bound states
and then evaluate the matrix elements of ${\cal H}_{I}$ between the degenerate many-body states.

The single-particle Andreev levels are found by solving Eq.~(\ref{h0}) subject to the appropriate matching conditions at $x=\pm L/2$.
For $\phi=\pi$ the energy eigenstates are Kramers pairs $\psi_{n,\sigma}$ indexed by $\sigma=\pm 1$, the eigenvalues of $\sigma_z$.
The energy eigenvalues $E_n$ satisfy
\begin{equation}
\tan \bar E_n \bar L = -\bar E_n (1-\bar E_n^2)^{-1/2}\,,
\end{equation}
where the bars denote dimensionless quantities $\bar E_n\equiv E_n/\Delta_0$ and $\bar L\equiv L/\xi=L\Delta_0/\hbar v_F$.
For $-\pi<2(\bar L - N\pi)\le \pi$ there are $N$ pairs of Andreev bound states in addition to the Majorana Kramers pair at $E_0=0$.
The wavefunctions $\psi_{n,\sigma}=(u_{n,\sigma},v_{n,\sigma})^T$ with $E_n \ge 0$ are
\begin{flalign}\label{uandv}
\!\begin{pmatrix} u_{n,\sigma}\\v_{n,\sigma}\end{pmatrix}\!
\!=\!{\cal A}_n e^{i\sigma\bar\mu \bar x - \sqrt{1-\bar E_n^2}|\bar x-\bar\ell(x)|}
\!\begin{pmatrix} (-1)^{n}e^{i\sigma \bar E_n \bar\ell(x)}\\-i\sigma e^{-i\sigma \bar E_n \bar\ell(x)}\end{pmatrix}\!,
\end{flalign}
where $\bar\ell(x)$ is $x/\xi$ for $|x|<L/2$ and $\sgn(x)\bar L/2$ for $|x|>L/2$.
The normalization factor satisfies
\begin{equation}
\mathcal{A}_n^{-2} = 2L+2\xi(1-\bar{E}_n^2)^{-1/2}.
\end{equation}
Particle-hole and time-reversal symmetries imply states with $E_{-n}=-E_{n}$ are related by
$\psi_{-n,\sigma}=-i\tau^y\psi_{n,\sigma}$ ($n>0$). The corresponding second quantized operators obey
$b_{-n,\sigma}=\sigma b_{n,-\sigma}^\dagger$ ($n>0$) and $b_{0,+} = i b_{0,-}^\dagger$.
It follows that the electron annihilation operator may be written as
\begin{equation}
c_\sigma(x) = u_{0,\sigma} b_{0,\sigma} + \sum_{n>0} u_{n,\sigma} b_{n,\sigma} - v_{n,\sigma}\sigma b^\dagger_{n,-\sigma}\,.
\label{csigma}
\end{equation}

We assume $\bar L > \pi/2$ so that there is at least one pair of excited bound states.
The four degenerate many-body states are $|\mu,\sigma\rangle = b_{1,\sigma}^\dagger |\mu\rangle_0$,
where $|\mu\rangle_0$ is the many-body ground state with $b_{n,\sigma}^\dagger b_{n,\sigma}=0$ ($n>0$) and $(-1)^{b_{0,+}^\dagger b_{0,+}} = \mu$.
For these four states $\mu=\pm 1$, the eigenvalues of $\mu^z$, distinguishes states with different fermion parity.
Under time reversal, $\Theta |\mu,\sigma\rangle = \sigma|-\mu,-\sigma\rangle$
and the operator $\Theta$ may be represented by $\Theta = \mu^x \sigma^y K$.
The most general interaction consistent with TRS and the conservation of fermion parity then has the form
$h_{I} = m_0+\vec m \cdot \vec\sigma\,\mu^z$. This splits the four states into two Kramers pairs with $E=m_0\pm|\vec m|$.

By plugging Eq.~(\ref{uandv}) and Eq.~(\ref{csigma}) into the density-density interactions~(\ref{interaction})
we find that $m_x=m_y=0$ and
\begin{equation}
m_z = \lambda \int_{-L/2}^{L/2} dx |u_{1,+}^*u_{0,-}-iu_{-1,-}u_{0,+}^*|^2\,,
\end{equation}
leading to the magnitude of level splitting
\begin{equation}
2m_z=\frac{\lambda}{\xi}
\bigg(\frac{1}{\sqrt{1-\bar{E}_1^2}}+\bar{L}\bigg)^{-1}.
\label{mz}
\end{equation}
When $\bar L\sim2.6$ the level splitting at $\phi=\pi$ can reach its largest amplitude $2m_z\sim 0.23\lambda/\xi$.
Physically, $\lambda=(e^2/\epsilon)\log(R_s/R)$, where $\epsilon$ is the dielectric constant;
$R$ and $R_s$ are respectively the penetration length and the screening radius of the edge states.
For $\epsilon=20$, $\xi=100$~nm, and $\log(R_s/R)=1$,
the splitting can reach $0.17$~meV, which is comparable to $\Delta_0$.
We also note that, in the presence of impurities, TRS allows scattering such as $(c_{\uparrow}^{\dagger}c_{\uparrow}c_{\downarrow}^{\dagger}\partial_{x}c_{\uparrow}-
c_{\downarrow}^{\dagger}c_{\downarrow}c_{\uparrow}^{\dagger}\partial_{x}c_{\downarrow})+\mbox{h.c.}$,
yielding to nonzero $m_{x,y}$. Estimating $m_{x,y}$ is beyond the scope here and we assume that they are in the same order of $m_z$.

Observing the fractional Josephson effect is complicated by scattering from thermally excited bulk quasiparticles,
which can cause the system to relax to the ground state before a cycle is completed.
It has been suggested that qualitative features of the $Z_2$ Josephson effect remain for equilibrium critical current measurements~\cite{alicea2}.   Here we consider a different method to demonstrate the $Z_4$ Josephson effect
by probing the phase dependence of the tunneling spectrum of Andreev bound states at the junction.

Consider a ring geometry where the phase difference $\phi$ across the junction is controlled by a weak applied magnetic field.
We propose tunneling into the junction region using an additional tunnel contact to probe the discrete many-body excitation spectrum.
This approach has been successfully used~\cite{tunneling,lesueur} to probe the Andreev spectra of 1D SNS junctions in similar geometries.
At low temperature, a weakly coupled tunnel junction probes the local tunneling density of states, $dI/dV \propto \rho(E=eV)$ with
\begin{equation}
\rho(E) = \sum_{N,\sigma} |\langle N|c_\sigma^\dagger|0\rangle|^2 \delta(E-E_N+E_0)\,,
\end{equation}
where $c_\sigma^\dagger$ is the creation operator for an electron with spin $\sigma$ in the junction
and $|N\rangle$ are the many-body states indicated in Fig.~\ref{fig1}(d).
Importantly, there is a selection rule, dictating that
the excited state $|N\rangle$ must have opposite fermion parity from the ground state $|0\rangle$.

\begin{figure}[t!]
\scalebox{0.42}{\includegraphics*{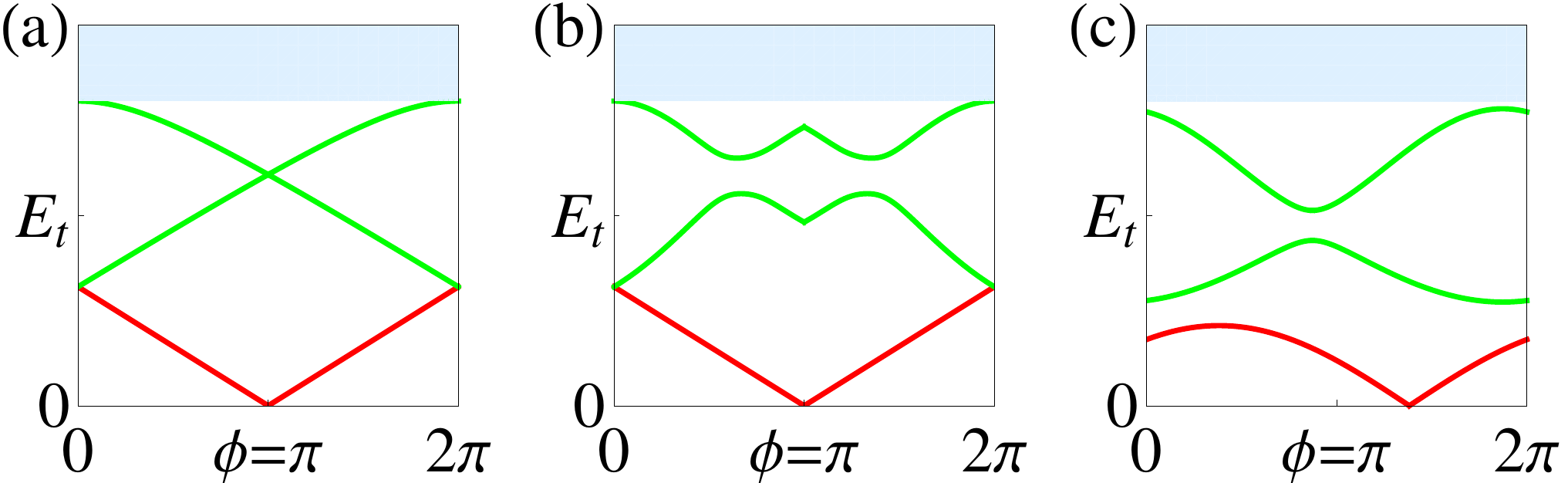}}
\caption{Phase dependence of the energies of peaks in the tunneling density of states.
In the non-interacting case (a) shows a degeneracy at $\phi=\pi$ that is lifted in (b) by interactions.
Breaking TRS lifts both degeneracies in (a), as shown in (c).}
\label{fig2}
\end{figure}

In Fig.~\ref{fig2} we plot the zero temperature peaks in the tunneling density of states based on the spectra in Fig.~\ref{fig1}.
$dI/dV$ must consist of peaks at $eV = E_N-E_0$, where $E_0$ is the ground state energy
and $E_N$ is the energy of a many-body excited state with opposite fermion parity.
Fig.~\ref{fig2}(a) shows the tunneling spectrum in the non-interacting electron approximation.
Fig.~\ref{fig2}(b) shows the spectrum when electron-electron interactions eliminate the many-body degeneracies.
Fig.~\ref{fig2}(c) shows the qualitative spectrum when TRS is strongly broken.
There are four important features: (i) Fig.~\ref{fig2}(a)-(c) all share a singularity in which the lowest peak goes to zero.
This is a consequence of the $Z_2$ Josephson effect, which requires a level crossing in the ground state when $\phi$ is advanced by $2\pi$.
Since this crossing changes the fermion parity it is visible in tunneling spectroscopy.
TRS fixes this singularity at $\phi=\pi$ whereas with broken TRS it can shift.
(ii) In the higher excited levels, similar singularities persist even with interactions
whereas with broken TRS they may disappear.
(iii) Fig.~\ref{fig2}(a) exhibits a degeneracy in the tunneling peaks at $\phi=\pi$,
which is split in Fig.~\ref{fig2}(b) by $2m_z$ in Eq.~(\ref{mz}) due to interactions.
(iv) Fig.~\ref{fig2}(a) and (b) both have the same Kramers degeneracy in the lowest two peaks at $\phi = 0$,
which is lifted in Fig.~\ref{fig2}(c) where TRS is broken.
Taken together, these four features would provide compelling evidences for the excitation spectrum responsible for the $Z_4$ Josephson effect.

The presence of a weak magnetic field that controls $\phi$ will break TRS and split the Kramers degeneracies.
The magnitude of the Zeeman splitting is $E_Z\sim g\mu_B B$.
Using $g\leq 1$ appropriate for the QSH edge states~\cite{theory,Ex-HgTe,Ex-Du}, we find $E_Z\leq 0.058\times$B[T]~meV,
which is negligible for $B<10$~mT relevant for the related experiments.

The $Z_2$ fractional Josephson effect can be understood in a weak coupling limit in which an electron---or half a Cooper pair---coherently tunnels between two Majorana bound states.
It is natural to ask whether there is a similar weak coupling version of the time-reversal-invariant $Z_4$ fractional Josephson effect.
For weak interactions, this is not possible because TRS prevents an energy gap from opening in the QSH edge state between the superconductors.
However, strong interactions can lead to an energy gap~\cite{wu,xu}.
We now show that in this case the Josephson effect is mediated by the tunneling of charge $e/2$ quasiparticles.
The domain wall at the interface between the gapped edge state and the superconductor behaves as a ``fractional'' Majorana mode, which is related to a $Z_4$ parafermion.

To describe the QSH edge state with strong interactions we adopt a bosonized representation
in which the electron creation operator is represented by $c_{R\uparrow (L\downarrow)}^\dagger \propto  e^{i(\varphi\pm\theta)}$,
where the bosonic variables satisfy $[\varphi(x),\theta(x')] = i \pi \Theta(x-x')$.
TRS forbids the single-particle backscattering term $\cos 2\theta$, which would lead to a magnetic energy gap.
However, the {\it pair} backscattering term $\cos 4\theta$ respects TRS and will be present---either as a momentum conserving process if $\mu = 0$ or as a impurity scattering process.
We thus consider the Hamiltonian ${\cal H} = {\cal H}_0 + {\cal H}_{I} + {\cal H}_\theta + {\cal H}_\varphi$ for the QSH edge state, where
\begin{equation}\label{h0I}
{\cal H}_0 +{\cal H}_{I} = \frac{v_F}{2\pi}\left[ (\partial_x\theta)^2 + (\partial_x\varphi)^2 \right] + \frac{\lambda(x)}{\pi^2}(\partial_x\theta)^2\,,
\end{equation}
with $\lambda(x) = \lambda \Theta(|x|-L/2)$. Eq.~(\ref{h0I}) is the Hamiltonian for a single channel Luttinger liquid,
with Eq.~(\ref{interaction}) being its interaction.
The superconducting proximity effect and the pair backscattering may be described by
\begin{flalign}
&{\cal H}_\varphi \!= u_0\big[\Theta(-\frac{L}{2}-x) \cos 2\varphi + \Theta(x-\frac{L}{2}) \cos (2\varphi- \phi)\big],\nonumber\\
&{\cal H}_\theta =  v_0\,\Theta(\frac{L}{2}-|x|) \cos 4\theta \,.
\label{clockmodel}
\end{flalign}

\begin{figure}[t!]
\scalebox{0.55}{\includegraphics*{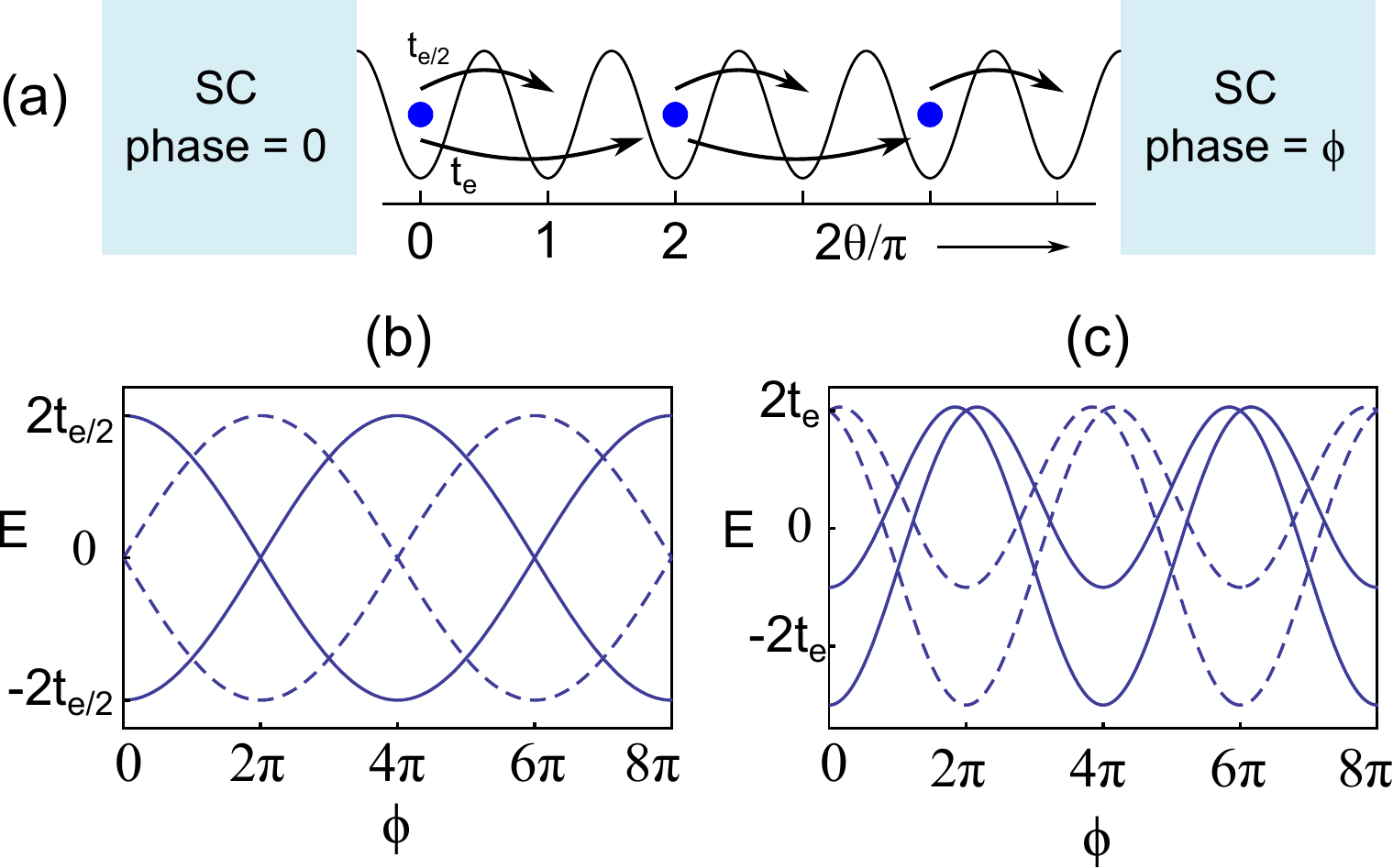}}
\caption{(a) Strong interactions pin the charge between the superconductors and lead to a fourfold ground state degeneracy.
Charge $e/2$ or charge $e$ tunneling processes lift the degeneracy, with an $8\pi$ periodicity in $\phi$, as shown in (b) for $t_e=0$ and (c) for $t_e = 2 t_{e/2}$.
Solid and dashed lines correspond to states with opposite fermion parity.}
\label{fig3}
\end{figure}

${\cal H}_\varphi$ introduces a superconducting energy gap into the QSH edge states coupled to the superconductors.
For weak interactions ${\cal H}_\theta$ is irrelevant, but ${\cal H}_\theta$ will flow to strong coupling and open a gap for $\lambda > 4.7 v_F$ for $\mu=0$ (corresponding to Luttinger parameter $g=(1+2\lambda/(\pi v_F))^{-1/2}<1/2$).
For impurity scattering, states are localized for $\lambda > 9.6 v_F$ ($g<3/8$)~\cite{giamarchi}.
It is simplest to analyze the limit of large $v_0$, where $\theta$ is pinned in the deep wells of the cosine potential, depicted in Fig.~\ref{fig3}(a).
This describes a magnetic state, in which the system spontaneously breaks TRS.
The presence of the $\cos 2\varphi$ introduces $2\pi$ jumps in $\theta$ that effectively makes $\theta$ an angular variable defined modulo $2\pi$,
reflecting the condensation of Cooper pairs.
There are thus four distinct minima of the $\cos 4\theta$ potential leading to a fourfold degenerate ground state for the junction.
When $v_0$ is finite, quantum tunneling between the minima will couple the four states, lifting their degeneracy with a characteristic pattern.
A tunneling event from $\theta = n\pi/2$ to $\theta = (n+1)\pi/2$ can be interpreted as the tunneling of a domain wall between the two degenerate magnetic states,
which is associated with a charge $e/2$~\cite{qi}. This has the effect of flipping the magnetization of the junction region while transferring a charge $e/2$.

If $|n\rangle$ denotes the state $\theta = n\pi/2$ (with $n$ defined modulo 4), the Hamiltonian in the degenerate subspace is
\begin{equation}
H = \sum_{n=1}^4 (-t_{e/2} e^{i\frac{\phi}{4}} |n\rangle\langle n+1|  - t_e e^{i\frac{\phi}{2}} |n\rangle\langle n+2| + \mbox{h.c.})\,,
\end{equation}
with an energy spectrum $E_{m=1,2,3,4}=-2t_{e/2}\cos[(\phi - 2\pi m)/4]-2t_e\cos[(\phi - 2\pi m)/2]$.
Here $t_{e/2}$ is the amplitude for tunneling a single $e/2$ quasiparticle whereas $t_e$ is the amplitude for tunneling charge $e$.
In general there will also be a contribution from tunneling charge $2e$ Cooper pairs across the junction,
which only gives an overall $\phi$ dependent shift to all four energy levels.
In Fig.~\ref{fig3}(a) and (b) we show $E_m(\phi)$ in the cases where tunnelings are dominated by $t_e=0$ and $t_e= 2 t_{e/2}$, respectively.
They share a pattern of fermion parity degeneracies (at $\phi = \pi$) and Kramers degeneracies (at $\phi = 0$)
that guarantee an $8\pi$ periodicity when $\phi$ is advanced adiabatically.

Eq.~(\ref{clockmodel}) is similar to models that have recently been introduced to describe ``fractional'' Majorana modes
in superconductor--fractional quantum Hall or fractional topological insulator structures~\cite{clarke,lindner,cheng,vaezi,mong,loss}.
These models share competing terms $\cos p\theta$ and $\cos 2\varphi$ with Eq.~(\ref{clockmodel}), which
are analogous to the order and disorder variables of a $Z_p$ clock model~\cite{fradkin}.
For $p=3$ the critical point of the $Z_3$ clock model is described by the $Z_3$ parafermion conformal field theory~\cite{zamolodchikov}.
In this case an interface between regions dominated by $\cos p\theta$ and by $\cos 2\varphi$ binds a local $Z_3$ parafermion,
which is related via a similar construction~\cite{teokane} to the quasiparticles of the Read-Rezayi state~\cite{readrezayi}.
The $p=4$ case of interest here is slightly different.
The $Z_4$ clock model and the $Z_4$ parafermion model are not equivalent,
but are rather two different points in the more general Ashkin-Teller model.
Nonetheless, the domain wall defines an excitation similar to a $Z_4$ parafermion,
and a pair of such defects encodes a fourfold degeneracy.
The domain walls that occur in superconductors coupled to fractionalized states with charge $e/m$ quasiparticles
involve a similar $Z_{2m}$ clock model (which also differs from the parafermion model), and lead to a $Z_{2m}$ fractional Josephson effect.
Despite the mathematical similarity, there is an important difference between Eq.~(\ref{clockmodel}) and the models based on
fractionalized states \cite{clarke,lindner,cheng,vaezi,mong,loss}.  In the later case,
the ground state degeneracy defined by a pair of domain walls is a {\it topological} degeneracy that can not be lifted by any local perturbation.
By contrast, half of the fourfold degeneracy defined by the Josephson junction here is a {\it local} degeneracy that can be lifted by a TRS breaking Zeeman field $h \cos 2\theta$.
In Fig.~\ref{fig3}, this eliminates the crossings between states with the same parity.
The fourfold degeneracy here, however, is a {\it symmetry-protected} degeneracy that is guaranteed as long as TRS is not violated.

\indent{\it Acknowledgement.}---We thank Jeffrey Teo for helpful discussions.
This work was supported by DARPA grant SPAWAR N66001-11-1-4110 and
a Simons Investigator award to CLK from the Simons Foundation.

\end{document}